# Application of Fractional Derivatives in Characterization of ECG graphs of Right Ventricular Hypertrophy Patients

**Srijan Sengupta[1a], UttamGhosh[1b], Susmita Sarkar[1c] and Shantanu Das [2]**
[1] Department of Applied Mathematics, University of Calcutta, Kolkata, India
[2] Reactor Control Systems Design Section E & I Group BARC Mumbai, India

Email:[1a]srijansengupta.math@gmail.com, [1b]uttam_math@yahoo.co.in, [1c]susmita62@yahoo.co.in, [2]shantanu@barc.gov.in

**Abstract**

There are many functions which are continuous everywhere but non-differentiable at someor all points such functions are termed as unreachable functions. Graphs representing suchunreachable functions are called unreachable graphs. For example ECG is such an unreachable graph. Classical calculus fails in their characterization as derivatives do not exist at the unreachable points. Such unreachable functions can be characterized by fractional calculus as fractional derivatives exist at those unreachable points where classical derivatives do not exist. Definition of fractional derivatives have been proposed by several mathematicians like Grunwald-Letinikov, Riemann-Liouville, Caputo and Jumarie to develop the theory of fractional calculus. In this paper we have used Jumarie type fractional derivative and consequently the phase transition (P.T.) which is the deference between left fractional derivative and right fractional derivatives to characterize those points. A comparative study has been done between normal ECG sample and problematic ECG sample (Right Ventricular Hypertrophy) by the help of above mentioned mathematical tool.

## 1. Introduction:

Basically fractional calculus was born in the year 1695 by raising a letter written by Leibniz to L'Hospital that "Can the meaning of derivatives with integer order be generalized to derivatives with non-integer orders?".It has become a rich area of research in the field of basic sciences and engineering since second half of the twentieth century[1-11]. This calculus is the generalization of differentiation and integration to an arbitrary order having applications in fluid mechanics, quantum mechanics,biomathematics including signal processing, control theory,etc.The most useful definitions of fractional derivative are Riemann-Liouvillie[2], Caputo[12], and Modified Riemann-Liouvillie [13,14]. Jumarie modified the Riemann-Liouvillie definition of fractional derivative to make derivative of constant zerowhich is non-zero in Riemann-Liouvillie sense[2,3]. Jumarie also modified the Caputo definition of fractional derivative as Caputo definition is applicable only for differentiable functions [2,3,12].

ECG(Electro-Cardiogram) is a continuous graph having non-differentiable points where classical derivative does not exist but fractional derivatives exist. Thus concept of fractional calculus thus can be applied to characterize those non-differentiable points of ECG graphs which may give a better measure for diagnosis of different heart diseases.

In this paper we shall characterize non-differentiable or unreachable points of ECG graphs using left modified and right modified Riemann-Liouvillie definitions[1-3,14] of half order fractional derivatives.The difference between left and right modified Riemann-Liouvillie derivative at a non differentiable point is known as Phase Transition (P.T.) at that point. In this paper we also find the mean and standard deviation of all non-differentiable points of ECG to get a better solution to interpret this type of ECGs. To characterize these graphs ECGs are taken from internet arbitrarily. ECG is the pictographic representation of electrical charge depolarization and repolarization of the heart muscle[16].There are several types of heart diseases such as right ventricular hypertrophy, left ventricular hypertrophy, right bundle branch block, left bundle branch block etc.[16] which can be detected by finding level of phase transition at some particular leads of patient's ECGs. Our main objective of this paper is to find some measures which will help the medical experts to diagnose right ventricular hypertrophy (RVH) disease from patient'sECG[16-17].

The rest of this paper is organized as follows: Section 2 describes some preliminaries about this paper; Section 3 describes about ECG and the right ventricular hypertrophy (RVH) disease in detail; Section 4 describes four theorems to obtain fractional derivative ,phase transition, mean and standard deviation of phase transition values at the non-differentiable points of any ECG which are found for abnormal and normal ECGs in tabular form in section 5 and 6 respectively.In Section 5 and 6 Doctoral view of that ECGs are also given in tabular form. Then section 7 we draw the bar diagrams of phase transition values of non-differentiable points of leads of different ECGs and  finally this paper is concluded in Section 8.

## 2.0 Some Preliminaries

### 2.1 Fractional Order Derivative

The fractional derivative is one of the oldest topic in mathematics but the definition of fractional derivative is modified day by day by many Mathematicians like Leibnitz, L-hospital, Grunwald-Letnikov, Riemann-Liouvillie, Caputo, Jumarie and others. Jumarie definition is basically modification of the Riemann-Liouvillie definition of fractional derivative as fractional order derivative of any constant is non zero in Riemann-Liouvillie sense whereas it is zero in Jumarie and Caputo sense. But the Caputo definition of fractional derivative can be applied only on differentiable functions whereas Jumarie definition of fractional derivative is applied for any continuous functions wherever it may be differentiable or not. In this paper we have used the Jumarie definition of fractional derivative to characterize the non-differentiable points in ECGs. The definition of fractional derivatives are as follows:

### 2.1.1 Grunwald-Letinikov definition

Let $f(x)$ be any function then the $\alpha$-th order derivative $\alpha \in \mathbb{R}$ of $f(x)$ is defined by

$$_aD_x^\alpha f(x) = \lim_{\substack{h \to 0 \\ nh \to t-a}} h^{-\alpha} \sum_{r=0}^{n} \binom{\alpha}{r} f(x-rh) = \frac{1}{\Gamma(-\alpha)} \int_a^x \frac{f(\xi)}{(x-\xi)^{\alpha+1}} d\xi \qquad (1)$$

Where $\alpha$ is any arbitrary number real or complex and $\binom{\alpha}{r} = \frac{\alpha!}{r!(\alpha-r)!} = \frac{\Gamma(\alpha+1)}{\Gamma(r+1)\Gamma(\alpha-r+1)}$

The above formula becomes fractional order integration if we replace $\alpha$ by $-\alpha$ which is

$$_aD_x^{-\alpha} f(x) = \frac{1}{\Gamma(\alpha)} \int_a^x (x-\xi)^{\alpha-1} f(\xi) d\xi \qquad (2)$$

Using the above formula we get for $f(x) = (x-a)^\gamma$,

$$_aD_x^\alpha (x-a)^\gamma = \frac{1}{\Gamma(-\alpha)} \int_a^x (x-\xi)^{-(\alpha+1)} (\xi-a)^\gamma d\xi$$

Using the substitution $\xi = a + \eta(x-a)$ we have for $\xi = a$ , $\eta = 0$ and for $\xi = x$, $\eta = 1$; $d\xi = (x-a)d\eta$ , $(x-\xi) = x-a-\eta(x-a) = (x-a)(1-\eta)$; $(\xi-a) = \eta(x-a)$, we get the following

$$_aD_x^\alpha(x-a)^\gamma = \frac{1}{\Gamma(-\alpha)}\int_a^x (x-\xi)^{-(\alpha+1)}(\xi-a)^\gamma d\xi = \frac{(x-a)^{\gamma-\alpha}}{\Gamma(-\alpha)}\int_0^1 \eta^\gamma(1-\eta)^{-(\alpha+1)}d\eta$$

$$= \frac{\Gamma(\gamma+1)}{\Gamma(\gamma+1-\alpha)}(x-a)^{\gamma-\alpha}, (\alpha<0, \gamma>-1)$$

If the function $f(x)$ be such that $f^k(x)$, $k=1,2,3,....,m+1$ is continuous in the closed interval $[a,x]$ and $m \leq \alpha < m+1$ then

$$_aD_x^\alpha f(x) = \sum_{k=0}^m \frac{f^{(k)}(a)(x-a)^{-\alpha+k}}{\Gamma(-\alpha+k+1)} + \frac{1}{\Gamma(-\alpha+m+1)}\int_a^x (x-\xi)^{m-\alpha}f^{(m+1)}(\xi)d\xi \qquad (3)$$

### 2.1.2 Riemann-Liouville (R-L) definition of fractional derivative

Let the function $f(x)$ is one time integrable then the integro-differential expression

$$_aD_x^\alpha f(x) = \frac{1}{\Gamma(-\alpha+m+1)}\left(\frac{d}{dx}\right)^{m+1}\int_a^x (x-\xi)^{m-\alpha}f(\xi)d\xi \qquad (4)$$

is known as the Riemann-Liouville definition of fractional derivative [3] with $m \leq \alpha < m+1$.

In Riemann-Liouville definition the function $f(x)$ is getting fractionally integrated and then differentiated $m+1$ whole-times but in obtaining the formula (3) $f(x)$ must be $m+1$ time differentiable. If the function $f(x)$ is $m+1$ whole times differentiable then the definition (1), (3) and (4) are equivalent.

Using integration by parts formula in (2), that is $_aD_x^{-\alpha}f(x) = \frac{1}{\Gamma(\alpha)}\int_a^x (x-\xi)^{\alpha-1}f(\xi)d\xi$ we get

$$_aD_x^{-\alpha}f(x) = \frac{f(a)(x-a)^\alpha}{\Gamma(\alpha+1)} + \frac{1}{\Gamma(\alpha+1)}\int_a^x (x-\xi)^\alpha f'(\xi)d\xi \qquad (5)$$

The left R-L fractional derivative is defined by

$$_aD_x^\alpha f(x) = \frac{1}{\Gamma(k-\alpha)}\left(\frac{d}{dx}\right)^k \int_a^x (x-\xi)^{k-\alpha-1}f(\xi)d\xi$$

$$k-1 \leq \alpha < k \qquad (6)$$

and the right R-L derivative is

$$_xD_b^\alpha f(x) = \frac{1}{\Gamma(k-\alpha)}\left(-\frac{d}{dx}\right)^k \int_x^b (\xi-x)^{k-\alpha-1}f(\xi)d\xi \qquad k-1 \leq \alpha < k \qquad (7)$$

In above definitions $k \in \mathbb{Z}$ i.e. integer just greater than alpha and $\alpha \succ 0$ $\alpha \in \mathbb{R}$

From the above definition it is clear that if at time $t=x$ the function $f(x)$ describes a certain dynamical system developing with time then for $\xi < x$, where $x$ is the present time then state $f(x)$ represent the past time and similarly if $\xi > x$ then $f(x)$ represent the future time. Therefore the left derivative represents the past state of the process and the right hand derivative represents the future stage.

### 2.1.3 Caputo definition of fractional derivative

In the R-L type definition the initial conditions contains the limit of R-L fractional derivative such as $\lim_{x \to a} {_aD_t^{\alpha-1}} = b_1$ etc that is fractional initial staes. But if the initial conditions are $f(a)=b_1, f'(a)=b_2...$

type then R-L definition fails and to overcome these problems M. Caputo [9] proposed new definition of fractional derivative in the following form

$$^C_a D^\alpha_t f(t) = \frac{1}{\Gamma(\alpha-n)} \int_a^t \frac{f^{(n)}(\tau)}{(t-\tau)^{\alpha+1-n}} d\tau, n-1 < \alpha < n \quad (8)$$

Under natural condition on the function $f(t)$ and as $\alpha \to n$ the Caputo derivative becomes a conventional *n*-th order derivative of the function. The main advantage of the Caputo derivative is the initial conditions of the fractional order derivatives are conventional derivative type-i.e requiring integer order states.

In R-L derivative the derivative of Constant (C) is non-zero. Since

$$_0 D^\alpha_t C = \frac{1}{\Gamma(-\alpha)} \int_0^t (t-\tau)^{-\alpha-1} C d\tau = \frac{Ct^{-\alpha}}{\Gamma(-\alpha)} \int_0^1 (1-x)^{-\alpha-1} dx, \text{Where } \tau = tx = -\frac{Ct^{-\alpha}}{\Gamma(-\alpha)} \frac{(1-x)^{-\alpha}}{-\alpha} \Big]_0^1$$

$$= \frac{Ct^{-\alpha}}{\Gamma(1-\alpha)}$$

which differ from the classical derivative.

### 2.1.4 Modified definitions of fractional derivative

On the other hand to overcome the misconception derivative of a constant is zero in the conventional integer order derivative Jumarie [6] revised the R-L derivative in the following form

$$D^\alpha_x f(x) = \frac{1}{\Gamma(-\alpha)} \int_0^x (x-\xi)^{-\alpha-1} f(\xi) d\xi, \text{for } \alpha < 0$$

$$= \frac{1}{\Gamma(1-\alpha)} \frac{d}{dx} \int_0^x (x-\xi)^{-\alpha} [f(\xi) - f(0)] d\xi, \text{for } 0 < \alpha < 1$$

$$= \left(f^{(\alpha-n)}(x)\right)^{(n)} \text{ for } n \leq \alpha < n+1, n \geq 1.$$

The above definition [6] is developed using left R-L derivative. Similarly using the right R-L derivative other type can be develop. Note in the above definition for negative fractional orders the expression is just Riemann-Liouvelli fractional integration. The modification is carried out in R-L the derivative formula, for the positive fractional orders alpha. The idea is to remove the offset value of function at start point of the fractional derivative from the function, and carry out R-L derivative usually done for the function.

Using the right R-L definition we established [8] the right modified definition of the fractional derivative in the following form

$$D^\alpha_x f(x) = -\frac{1}{\Gamma(-\alpha)} \int_x^b (\xi-x)^{-\alpha-1} f(\xi) d\xi, \quad \text{for } \alpha < 0$$

$$= -\frac{1}{\Gamma(1-\alpha)} \frac{d}{dx} \int_x^b (\xi-x)^{-\alpha} [f(b) - f(\xi)] d\xi, \quad \text{for } 0 < \alpha < 1$$

$$= \left(f^{(\alpha-n)}(x)\right)^{(n)} \text{ for } n \leq \alpha < n+1, n \geq 1.$$

### 2.2 Unreachable Functions

Those functions which are continuous everywhere but not-differentiable at some point or all point, termed as unreachable functions like as (i) |x-a| is unreachable at x=a where a is any point in real line, (ii)

$x \sin \dfrac{1}{x}$ is unreachable at x=0 etc. similarly Weierstrass function, white noise, Brownianmotion[1] are also unreachable functions.ECGs are also these types of graphs which are continuous but have some non-differentiable points. The points Q,R,S of QRS complexes of a PQRST wave of any lead of ECG may be differentiable or non-differentiablewhich is shown in fig-1. For non-differentiable points we cannot describe it by classical sense but here fractional sense is applicable.

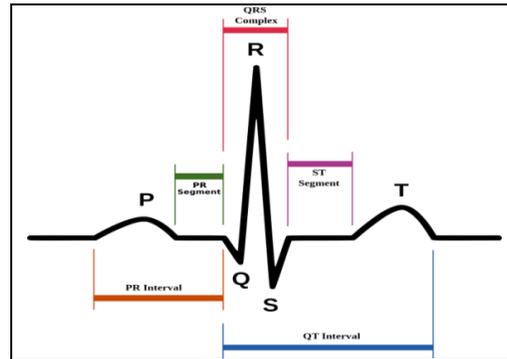

**Fig-1: A normal shape of PQRST wave in ECG**

### 2.3 Phase Transition
The difference between the left and right modified Riemann-Liouvillie derivative at unreachable points of any continuous graph is termed as Phase Transition at that point of that graph[14].

### 3.0 About ECG

### 3.1 Some Preliminaries on ECGLeads
The heart's electrical activity recorded from electrodes on the body surface by the 12-lead electrocardiogram. This section describes the basic components of the ECG and the lead system used to record the ECG tracings. There are twelve leads which are I, II, III, AVR, AVL, AVF obtained from the 'limb' and V1,V2, V3,V4, V5, V6 which are obtained from the 'chest'. The first six represents the different electrical situation of the Atria and the later six represents the different electrical situation of the Ventricle. The six V leads (V1-V6) look at the heart in horizontal plane, from the front and the left side [16].

I, II, and AVL-looks at the left lateral surface of the heart.
III and AVF-looks at the interior surface of the atria.
AVR looks at the right atria.
V1 and V2 look at the right ventricle.
V3 and V4 look at septum between ventricles and the anterior wall of the left ventricle.
V5 and V6 represent the anterior and lateral wall of the left ventricle

### 3.2 Right Ventricular Hypertrophy (RVH)
Right Ventricular Hypertrophy is a condition where the muscle wall of heart's right ventricle becomes thickened (hypertrophy). This can arise due to chronic pressure overload, similar to that of Left Ventricular Hypertrophy[18]. Since blood travels through the right ventricle to the lungs via pulmonary arteries,if the condition of RVH occurs then pulmonary circulation decreases i.e. blood does not flow well from the heart to the lungs of the body, then extra stress can be placed on the right ventricle.An ECG with Right Ventricular Hypertrophy may or may not show a right axis deviation on the graph[16].

**Symptoms:**

1. Pulmonary Hypertension
2. Pulmonary embolism
3. Chronic lungs disease
4. Mitral stenosis
5. Congenital heart disease

**Criteria of RVH in ECG lead:**
1. R wave in V1 lead + S wave in V5 and V6 lead is greater than 11 mm
2. R wave in V1 lead is greater than 7 mm
3. The ratio of R wave and S wave in V1 lead is greater than 1 mm
4. R wave in V5 or V6 lead is less than 5 mm
5. S wave in V5 or V6 lead is greater than 7 mm
6. The ratio of R wave and S wave in V5 or V6 lead is less than 1 mm

**4. Application of fractional derivative in ECG Graph:**
Now we have to study the non-differentiable points of the PQRST waves in ECG leads with the help of fractional derivatives. Herewe have to find out the half-order fractional derivative (both left and right) and calculate the corresponding Normalized Phase Transition values(N.P.T.) by the help of the following theorems.

**Theorem 1:** Let us consider the function
$$f(x) = \begin{cases} ax+b, & p \leq x \leq q \\ cx+d, & q \leq x \leq r \end{cases}$$

with $a \neq c$. It is continuous at $x=q$ such that $aq+b = cq+d$ but not differentiable at that point. Then left fractional derivative, right fractional derivative and phase transition at that point x=q are respectively:

$$f_L^{(\alpha)}(q) = \frac{a(q-p)^{1-\alpha}}{\Gamma(2-\alpha)} \qquad f_R^{(\alpha)}(q) = \frac{c(r-q)^{1-\alpha}}{\Gamma(2-\alpha)};$$

$$P.T = \frac{\left(a(q-p)^{1-\alpha} - c(r-q)^{1-\alpha}\right)}{\Gamma(2-\alpha)}$$

**Theorem 2:** Let us consider the function
$$f(x) = \begin{cases} ax+b, & p \leq x \leq q \\ cx^2 + gx + h, & q \leq x \leq r \end{cases}$$

with $a \neq 2cq+g$. It is continuous at $x=q$ such that $aq+b = cq^2+gq+h$ but not differentiable at that point. Then left fractional derivative, right fractional derivative and corresponding phase transitionat that point x=qare respectively:

$$f_L^{(\alpha)}(q) = \frac{a(q-p)^{1-\alpha}}{\Gamma(2-\alpha)}$$

$$f_R^{(\alpha)}(q) = \frac{1}{\Gamma(1-\alpha)}\left(c\alpha\frac{(r-q)^{2-\alpha}}{2-\alpha} + \left(\frac{\alpha(2cq+g)}{1-\alpha} + (cq+cr+g)\right)(r-q)^{1-\alpha}\right)$$

$$P.T. = \frac{1}{\Gamma(1-\alpha)}\left(\frac{a(q-p)^{1-\alpha}}{1-\alpha} - c\alpha\frac{(r-q)^{2-\alpha}}{2-\alpha} - \left(\frac{\alpha(2cq+g)}{1-\alpha} + (cq+cr+g)\right)(r-q)^{1-\alpha}\right)$$

**Theorem 3**: Let us consider the function

$$f(x) = \begin{cases} ax^2+bx+c, & p \leq x \leq q \\ gx+h, & q \leq x \leq r \end{cases}$$

with 2aq+b≠g. It is continuous at x=q such that $aq^2+bq+c = gq+h$ but not differentiable at that point. Then left fractional derivative, right fractional derivative and corresponding phase transition at that point x=q are respectively:

$$f_L^{(\alpha)}(q) = \frac{1}{\Gamma(1-\alpha)}\left(-a\alpha\frac{(q-p)^{2-\alpha}}{2-\alpha} + \left(\frac{\alpha(2aq+b)}{1-\alpha} + (ap+aq+b)\right)(q-p)^{1-\alpha}\right)$$

$$f_R^{(\alpha)}(q) = \frac{g(r-q)^{1-\alpha}}{\Gamma(2-\alpha)}$$

$$P.T. = \frac{1}{\Gamma(1-\alpha)}\left(-a\alpha\frac{(q-p)^{2-\alpha}}{2-\alpha} + \left(\frac{\alpha(2aq+b)}{1-\alpha} + (ap+aq+b)\right)(q-p)^{1-\alpha} - \frac{g(r-q)^{1-\alpha}}{1-\alpha}\right)$$

**Theorem 4**: Let us consider the function

$$f(x) = \begin{cases} ax^2+bx+c, & p \leq x \leq q \\ gx^2+hx+m, & q \leq x \leq r \end{cases}$$

with a≠g or b≠h. It is continuous at x=q such that $aq^2+bq+c = gq^2+hq+m$ but not differentiable at that point. Then left fractional derivative, right fractional derivative and corresponding phase transition at that point x=q are respectively:

$$f_L^{(\alpha)}(q) = \frac{1}{\Gamma(1-\alpha)}\left(-a\alpha\frac{(q-p)^{2-\alpha}}{2-\alpha} + \left(\frac{\alpha(2aq+b)}{1-\alpha} + (ap+aq+b)\right)(q-p)^{1-\alpha}\right)$$

$$f_R^{(\alpha)}(q) = \frac{1}{\Gamma(1-\alpha)}\left(g\alpha\frac{(r-q)^{2-\alpha}}{2-\alpha} + \left(\frac{\alpha(2gq+h)}{1-\alpha} + (gr+gq+h)\right)(r-q)^{1-\alpha}\right)$$

$$P.T. = \frac{1}{\Gamma(1-\alpha)}\left(\begin{array}{l}-a\alpha\frac{(q-p)^{2-\alpha}}{2-\alpha} + \left(\frac{\alpha(2aq+b)}{1-\alpha} + (ap+aq+b)\right)(q-p)^{1-\alpha} - \\ g\alpha\frac{(r-q)^{2-\alpha}}{2-\alpha} - \left(\frac{\alpha(2gq+h)}{1-\alpha} + (gr+gq+h)\right)(r-q)^{1-\alpha}\end{array}\right)$$

## 5. Normal ECG

Now we shall characterize the ECG graphs by the help of fractional derivative to compare normal ECGs with abnormal ECGs (RVH). For this purpose we shall consider ½- order fractional derivatives. The non-differentiable points Q, R, S of QRS complexes of PQRST wave of any leads are used here usually points, not as waves. If Q or S point is not prominent at QRS complex of any lead of the ECGs under consideration then we cannot find the Left Fractional Derivative and Right Fractional Derivative at that point. We have denoted those cases by 'NA' i.e. 'Not Arise'. To investigate the characteristics of the ECG we here consider normal ECG and problematic ECGs(in our case RVH ECG to be compared with normal ECG).

Since RVH is characterized by ratio of R and S wave in V1, V5 and V6 leads. Thus these paper contributed only P.T. values at non-differentiable points at those leads. So our concern is to find any distinguishing measurements of P.T values at non-differentiable points on those leads to characterize the problematic ECG (in our case RVH) with normal ECG.

Thus our paper contributed only P.T. values at non-differentiable points at those leads for RVH ECGs. So our concern to find any distinguishing measurements of P.T. values at non-differentiable points R and S wave in V1 ,V5 and V6 leads to characterize the problematic ECG (in our case RVH) with normal ECG. Following tables are new and constructed from our fractional calculus methodology.

i) Left and right fractional derivatives and phase transition have been calculated at the Q,R,S points of V1 lead in table, V5 lead in table  and V6 lead in table . Then we also find mean and standard deviation of P.T. values at non-differentiable points of considerable leads of ECG graphs. In these tables we have not calculated fractional derivatives and hence phase transition at those points where integer order derivatives exist as the curves are smooth there and fractional derivatives are Not Arise (NA) there. Since RVH is described by ratio of R:S in V1,V5 and V6 leads.

### i. Normal ECG1:

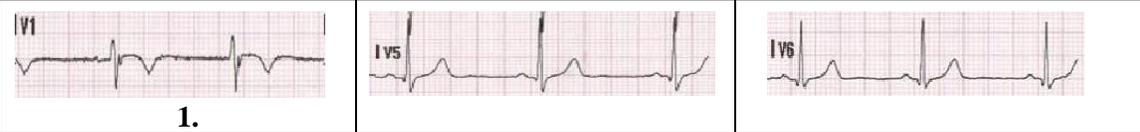

1.

| PQRST waves | Length of R wave (mm) in V1 | Length of S wave (mm) in V1 | Length of S wave (mm) in V5 | Length of S wave (mm) in V6 | Length of R wave (mm) in V5 | Length of R wave (mm) in V6 | R:S in V1(mm) | R:S in V5(mm) | R:S in V6(mm) |
|---|---|---|---|---|---|---|---|---|---|
| 1st | 4 | 9.5 | 21 | 15 | 18 | 14.5 | 0.42 | 0.86 | 0.97 |
| 2nd | 4.5 | 10.5 | 21 | 15 | 18 | 14.5 | 0.43 | 0.86 | 0.97 |
| 3rd |  |  | 20.5 | 14 | 18 | 13.5 |  | 0.88 | 0.96 |

| Length of R wave in V1+S wave in V5 (mm) S1 | Length of R wave in V1+S wave in V6 (mm) S2 |
|---|---|
| 25 | 19 |
| 25.5 | 19.5 |
| 25 | 19 |
| 25.5 | 19.5 |
| 24.5 | 18 |
| 25 | 18.5 |

**Table 1A:** Length of S wave in V1, V2 and length of R wave in V5 and V6 and their ratios and length of S1, S2.

Here, we see that in table-1A lengths of R wave in V1 is less than 7 mm and length of R wave in V5 and V6 lead are greater than 5 mm but length of S wave in V5,V6 respectively are not less than 7 mm whereas ratios of R and S wave in V1 is less than 1 and in V5 and V6 are close to 1. So from Doctor's point of view this graph is normal ECG graph.

Now we have to consider above techniques mentioning in section5 characterizing this normal ECG first by us with Fractional Calculus. To do this job first we calculated Left Fractional Derivative and Right Fractional Derivative and corresponding Phase Transition (P.T.) values at non-differentiable points of considerable leads of examined ECG graphs; with these set of P.T. values we also determined mean and standard deviation of P.T. values for all non-differentiable points of considerable leads of examined ECG graphs to get my desired results.

In the following tables we have presented the fractional derivatives and phase transition values of the non-differentiable points of different leads of ECG presented in table-1A.

|  |  | Left Fractional Derivative | Right Fractional Derivative | Phase Transition |
|---|---|---|---|---|
| 1$^{ND}$ | For PQ:QR at Q | 0 | 6.383076 | 6.383076 |
|  | For QR:RS at R | 6.383076 | -15.159807 | 21.542883 |
|  | For RS:ST at S | -15.159807 | 10.372499 | 25.532306 |
| 2$^{ST}$ | For PQ:QR at Q | 0 | 5.077706 | 5.077706 |
|  | For QR:RS at R | 5.077706 | -16.755576 | 21.833282 |
|  | For RS:ST at S | -16.755576 | 7.978846 | 24.734422 |

**Table 1B:** Phase transition at the non-differentiable points Q,R,S of lead V1

|  |  | Left Fractional Derivative | Right Fractional Derivative | Phase Transition |
|---|---|---|---|---|
| 1$^{ST}$ | For PQ:QR at Q | -1.520174 | 28.723844 | 30.244018 |
|  | For QR:RS at R | 28.723844 | -33.511152 | 62.234996 |
|  | For RS:ST at S | -33.511152 | 3.385138 | 36.896290 |
| 2$^{ND}$ | For PQ:QR at Q | -1.595769 | 28.723844 | 30.319613 |
|  | For QR:RS at R | 28.723844 | -23.695962 | 52.419806 |
|  | For RS:ST at S | -23.695962 | 4.787307 | 28.483269 |
| 3$^{RD}$ | For PQ:QR at Q | -3.191538 | 28.723844 | 31.915382 |
|  | For QR:RS at R | 28.723844 | -32.713267 | 61.437111 |
|  | For RS:ST at S | -32.713267 | 2.820948 | 35.534215 |

**Table 1C:** Phase transition at the non-differentiable points Q,R,S of lead V5

|  |  | Left Fractional Derivative | Right Fractional Derivative | Phase Transition |
|---|---|---|---|---|
| 1$^{ST}$ | For PQ:QR at Q | -2.350790 | 23.138652 | 25.489442 |
|  | For QR:RS at R | 23.138652 | -16.925688 | 40.064340 |
|  | For RS:ST at S | -16.925688 | 3.191538 | 20.117226 |
| 2$^{ND}$ | For PQ:QR at Q | -2.350790 | 18.892630 | 21.243420 |
|  | For QR:RS at R | 18.892630 | -23.936537 | 42.829167 |
|  | For RS:ST at S | -23.936537 | 1.302940 | 25.239477 |
| 3$^{RD}$ | For PQ:QR at Q | -2.280261 | 21.542883 | 23.823144 |
|  | For QR:RS at R | 21.542883 | -15.797308 | 37.340191 |

|  | For RS:ST at S | -15.797308 | 2.393654 | 18.190962 |
|---|---|---|---|---|

**Table 1D:** Phase transition at the non-differentiable points Q,R,S of lead V6

Now we construct a table with mean and standard deviation of phase transition values at the non-differentiable points of considerable ECG graph to get a better solution.

| Non-differentiable Points | Examined Leads | Mean | SD |
|---|---|---|---|
| **Q point** | V1 lead | 5.730391 | 0.923035979 |
|  | V5 lead | 30.82633767 | 0.943897145 |
|  | V6 lead | 23.51866867 | 2.139323404 |
| **R point** | V1 lead | 21.6880825 | 0.205343102 |
|  | V5 lead | 58.69730433 | 5.451091094 |
|  | V6 lead | 40.07789933 | 2.744513121 |
| **S point** | V1 lead | 25.133364 | 0.564189187 |
|  | V5 lead | 33.63792467 | 4.515713494 |
|  | V6 lead | 21.182555 | 3.643018712 |

**Table 1E:** Mean and Standard Deviation of Phase transition at the non-differentiable points Q,R,S of lead V1,V5&V6

### ii. Normal ECG 2:

| PQRST waves | Length of R wave (mm) in V1 | Length of S wave (mm) in V1 | Length of S wave (mm) in V5 | Length of S wave (mm) in V6 | Length of R wave (mm) in V5 | Length of R wave (mm) in V6 | R:S in V1( mm) | R:S in V5( mm) | R:S in V6(mm) |
|---|---|---|---|---|---|---|---|---|---|
| 1st | 2.8 | 19.7 | 17.6 | 18 | 17.9 | 19.2 | 0.14 | 1.02 | 1.07 |
| 2nd | 3.1 | 18.8 | 19.6 | 18.2 | 19.4 | 19.5 | 0.16 | 0.99 | 1.07 |
| 3rd | 3.2 | 19.5 | 18.3 | 18.6 | 18.6 | 19.9 | 0.16 | 1.02 | 1.07 |

| Length of R wave in V1+S wave in V5 (mm) S1 | Length of R wave in V1+S wave in V6 (mm) S2 |
|---|---|
| 20.4 | 20.8 |
| 22.4 | 21 |
| 21.1 | 21.4 |
| 20.7 | 21.1 |
| 22.7 | 21.3 |
| 21.4 | 21.7 |
| 20.8 | 21.2 |
| 22.8 | 21.4 |
| 21.5 | 21.8 |

**Table 2A:** Length of S wave in V1, V2 and length of R wave in V5 and V6 and their ratios and length of S1, S2.

Here, we see that in table-2A lengths of R wave in V1 is less than 7 mm and length of R wave in V5 and V6 lead are greater than 5 mm but length of S wave in V5,V6 respectively are not less than 7 mm whereas ratios of R and S wave in V1 is less than 1 and in V5 and V6 are greater than 1. So from Doctor's point of view this graph is normal ECG graph.

In the following tables we have presented the fractional derivatives and phase transition values of the non-differentiable points of different leads of ECG presented in table-2A in similar way.

|  |  | P.T. values of V1 | P.T. values of V5 | P.T. values of V6 |
|---|---|---|---|---|
| 1st | For PQ:QR at Q | NA* | 23.4741 | 27.7903 |
|  | For QR:RS at R | 33.7053 | 53.9826 | 62.0691 |
|  | For RS:ST at S | 38.4275 | 32.0749 | NA |
| 2nd | For PQ:QR at Q | NA | 30.0893 | 27.4551 |
|  | For QR:RS at R | 34.3529 | 60.7762 | 63.7091 |
|  | For RS:ST at S | 36.2674 | 34.6120 | NA |
| 3rd | For PQ:QR at Q | NA | 24.4952 | 30.2656 |
|  | For QR:RS at R | 31.8500 | 50.1233 | 51.7140 |
|  | For RS:ST at S | 38.4311 | 26.6582 | NA |

**Table 2B:** Phase transition at the non-differentiable points Q,R,S of V1,V5 and V6 leads

*NA-If Q or S point is not prominent at QRS complex of any lead of the ECGs under consideration i.e. the curves are smooth at that point then we cannot find the Left Fractional Derivative and Right Fractional Derivative at that point. We have denoted those cases by 'NA' i.e. 'Not Arise'. The above table has this type of consideration.

Now we construct a table with mean and standard deviation of phase transition values at the non-differentiable points of considerable ECG graph to get a better solution.

| **Non-differentiable Points** | **Examined Leads** | **Mean** | **SD** |
|---|---|---|---|
| **Q point** | V1 lead | NA | NA |
|  | V5 lead | 26.01953 | 3.561308 |
|  | V6 lead | 28.50367 | 1.535056 |
| **R point** | V1 lead | 33.30273 | 1.299104 |
|  | V5 lead | 54.9607 | 5.393383 |
|  | V6 lead | 59.16407 | 6.503847 |
| **S point** | V1 lead | 37.70867 | 1.248175 |
|  | V5 lead | 31.11503 | 4.062849 |
|  | V6 lead | NA | NA |

**Table 2C:** Mean and Standard Deviation of Phase transition at the non-differentiable points Q,R,S of lead V1,V5&V6

6. **Problematic ECG**

**Problematic ECG1**

| PQRST waves | Length of R wave (mm) | Length of S wave (mm) | Length of S wave (mm) | Length of S wave (mm) | Length of R wave (mm) in | Length of R wave (mm) | R:S in V1(mm) | R:S in V5(mm) | R:S in V6(mm) |
|---|---|---|---|---|---|---|---|---|---|

| | in V1 | in V1 | in V5 | in V6 | V5 | in V6 | | | |
|---|---|---|---|---|---|---|---|---|---|
| 1st | 10.7 | 8.1 | 15.2 | 10.6 | 7.8 | 6.5 | 1.32 | 0.51 | 0.61 |
| 2nd | 10 | 8 | 17.8 | 11.1 | 8.8 | 7.1 | 1.25 | 0.49 | 0.64 |
| Length of R wave in V1+S wave in V5 (mm) S1 | | | | | Length of R wave in V1+S wave in V6 (mm) S2 | | | | |
| 25.9 | | | | | 21.3 | | | | |
| 28.5 | | | | | 21.8 | | | | |
| 25.2 | | | | | 20.6 | | | | |
| 27.8 | | | | | 21.1 | | | | |

**Table 3A:** Length of S wave in V1, V2 and length of R wave in V5 and V6 and their ratios andlength of S1, S2.

Here, we see that in table-3A lengths of R wave in V1 and length of S wave in V5,V6 respectively are all greater than 7 mm but length of R wave in V5 and V6 lead are not less than 5 mm whereas ratios of R and S wave in V1 is greater than 1 and in V5 and V6 are less than 1. Also length of S1, S2 are all greater than 11 mm of that table. So from Doctor's point of view this patient has cardiac problem which called Right Ventricular Hypertrophy.

In the following tables we have presented the fractional derivatives and phase transition values of the non-differentiable points of different leads of ECG presented in table-3A in similar way.

| | | P.T. values of V1 | P.T. values of V5 | P.T. values of V6 |
|---|---|---|---|---|
| 1st | For PQ:QR at Q | 26.0803932 | NA | NA |
| | For QR:RS at R | 36.4947865 | 33.2203698 | 25.1153 |
| | For RS:ST at S | 14.7037186 | 25.7151588 | 23.6948 |
| 2nd | For PQ:QR at Q | 17.4327977 | NA | NA |
| | For QR:RS at R | 26.8886555 | 39.3007419 | 20.8981 |
| | For RS:ST at S | 16.333122 | 35.4445989 | 20.3390 |

**Table 3B:** Phase transition at the non-differentiable points Q,R,S of V1,V5 and V6 leads

Now we construct a table with mean and standard deviation of phase transition values at the non-differentiable points of considerable ECG graph to get a better solution.

| Non-differentiable Points | Examined Leads | Mean | SD |
|---|---|---|---|
| **Q point** | V1 lead | 21.7566 | 6.114773 |
| | V5 lead | NA | NA |
| | V6 lead | NA | NA |
| **R point** | V1 lead | 31.69172 | 6.79256 |
| | V5 lead | 36.26056 | 4.299472 |
| | V6 lead | 23.0067 | 2.982054 |
| **S point** | V1 lead | 15.51842 | 1.152162 |
| | V5 lead | 30.57988 | 6.879753 |
| | V6 lead | 22.01692 | 2.372916 |

**Table 3C:** Mean and Standard Deviation of Phase transition at the non-differentiable points Q,R,S of lead V1, V5&V6

i. **Problematic ECG2:**

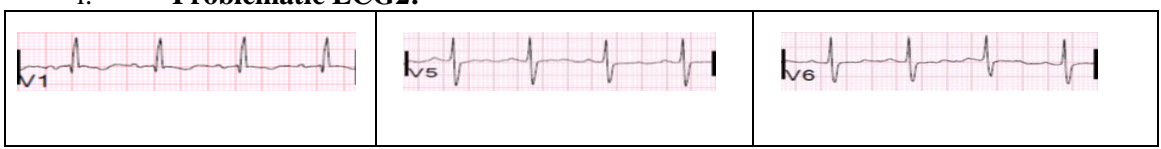

| PQRST waves | Length of R wave (mm) in V1 | Length of S wave (mm) in V1 | Length of S wave (mm) in V5 | Length of S wave (mm) in V6 | Length of R wave (mm) in V5 | Length of R wave (mm) in V6 | R:S in V1(mm) | R:S in V5(mm) | R:S in V6(mm) |
|---|---|---|---|---|---|---|---|---|---|
| 1st | 10.7 | 8.8 | 19.2 | 15.6 | 9.1 | 8.6 | 1.22 | 0.47 | 0.55 |
| 2nd | 10.1 | 9.3 | 19.4 | 15.3 | 10.3 | 8.8 | 1.09 | 0.53 | 0.58 |
|  | 11 | 9 | 17.1 | 14 | 9.1 | 8.1 | 1.22 | 0.53 | 0.58 |
|  | 10.9 | 8.9 | 19.8 | 15.2 | 10 | 8.8 | 1.22 | 0.51 | 0.58 |

| Length of R wave in V1+S wave in V5 (mm) S1 | Length of R wave in V1+S wave in V6 (mm) S2 |
|---|---|
| 29.9 | 26.3 |
| 30.1 | 26 |
| 27.8 | 24.7 |
| 30.5 | 25.9 |
| 29.3 | 25.7 |
| 29.5 | 25.4 |
| 27.2 | 24.1 |
| 29.9 | 25.3 |
| 30.2 | 26.6 |
| 30.4 | 26.3 |
| 28.1 | 25 |
| 30.8 | 26.2 |
| 30.1 | 26.5 |
| 30.3 | 26.2 |
| 28 | 24.9 |
| 30.7 | 26.1 |

**Table 4A:** Length of S wave in V1, V2 and length of R wave in V5 and V6 and their ratios and length of S1, S2.

Here, we see that in table-5A lengths of R wave in V1 and length of S wave in V5,V6 respectively are all greater than 7 mm but length of R wave in V5 and V6 lead are not less than 5 mm whereas ratios of R and S wave in V1 is greater than 1 and in V5 and V6 are less than 1. Also length of S1, S2 are all greater than 11 mm of that table. So from Doctor's point of view this patient has cardiac problem which called Right Ventricular Hypertrophy.

In the following tables we have presented the fractional derivatives and phase transition values of the non-differentiable points of different leads of ECG presented in table-4A in similar way.

|  |  | P.T. values of V1 | P.T. values of V5 | P.T. values of V6 |
|---|---|---|---|---|
| 1st | For PQ:QR at Q | 21.210452 | NA | NA |
|  | For QR:RS at R | 30.1310726 | 37.1478021 | 31.2891687 |
|  | For RS:ST at S | 15.7002922 | 34.5669156 | 25.1595267 |
| 2nd | For PQ:QR at Q | 16.7788313 | NA | NA |
|  | For QR:RS at R | 24.4743855 | 44.5219585 | 38.9823975 |
|  | For RS:ST at S | 13.7510722 | 36.6391728 | 21.4870607 |

|  | For PQ:QR at Q | 19.5922919 | NA | NA |
|---|---|---|---|---|
| 3rd | For QR:RS at R | 32.0811607 | 34.5425282 | 24.6282722 |
|  | For RS:ST at S | 16.05711705 | 25.9435303 | 22.9338048 |
| 4th | For PQ:QR at Q | 19.4358294 | NA | NA |
|  | For QR:RS at R | 20.4990607 | 47.6872047 | 39.3608674 |
|  | For RS:ST at S | 8.199727814 | 30.6882091 | 21.7925225 |

**Table 4B:** Phase transition at the non-differentiable points Q,R,S of V1,V5 and V6 leads

Now we construct a table with mean and standard deviation of phase transition values at the non-differentiable points of considerable ECG graph to get a better solution.

| Non-differentiable Points | Examined Leads | Mean | SD |
|---|---|---|---|
| **Q point** | V1 lead | 19.25435 | 1.834999 |
|  | V5 lead | NA | NA |
|  | V6 lead | NA | NA |
| **R point** | V1 lead | 26.79642 | 5.294556 |
|  | V5 lead | 40.97487 | 6.155181 |
|  | V6 lead | 33.56518 | 7.023408 |
| **S point** | V1 lead | 13.42705 | 3.629269 |
|  | V5 lead | 31.95946 | 4.708362 |
|  | V6 lead | 22.84323 | 1.664994 |

**Table 4c:** Mean and Standard Deviation of Phase transition at the non-differentiable points Q,R,S of lead V1,V5&V6

## 7. Bar diagram of Phase transition values

In this section we have drawn bar diagrams of phase transition values of non-differential points of V1,V5and V6 leads of the above described normal and problematic ECGs. Figure 1 represents the bar diagram for normal ECG and figure 2 represents the bar diagram for problematic ECGs.

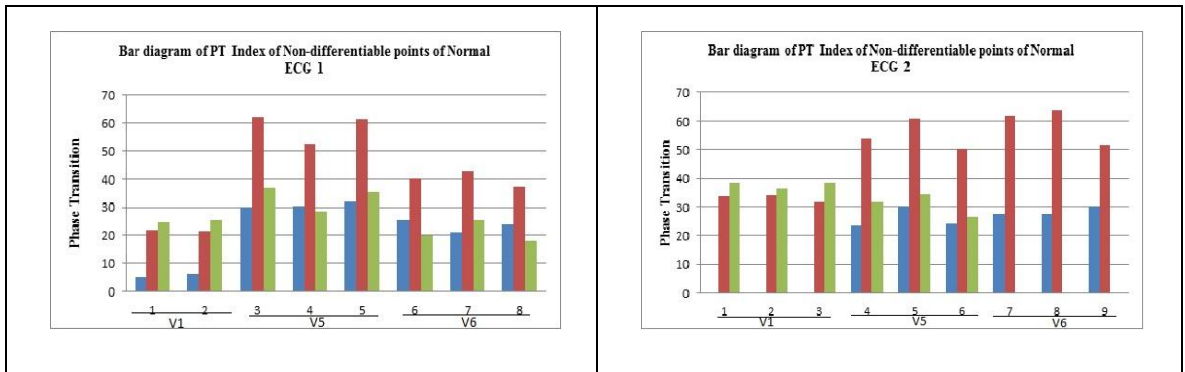

Figure1: Bar diagram of P.T. values of two Normal ECGS

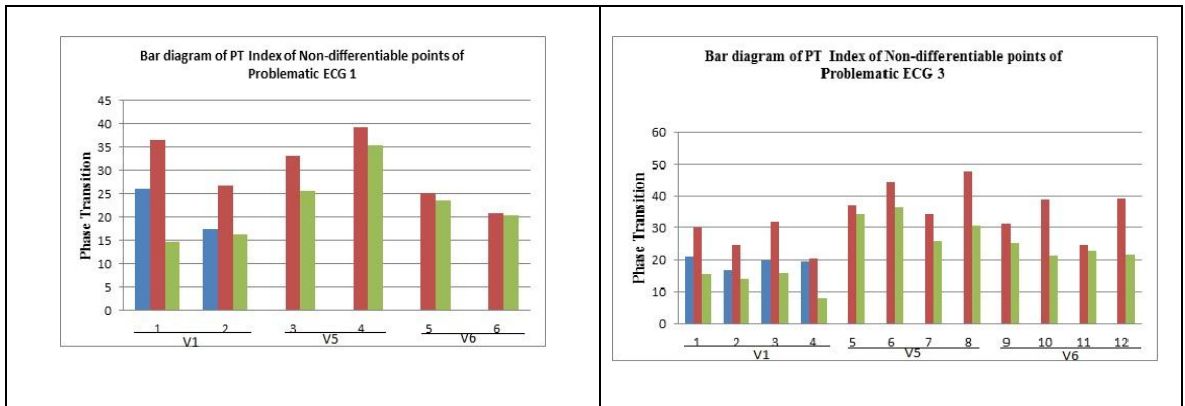
Figure2: Bar diagram of P.T. values of five Problematic ECGS

## 8. Conclusion

In this paper we have compared normal ECGs with the ECGs of RVH patients using fractional calculus. The tables represent the fractional derivatives of half-order together with their level of phase transition value at the non-differentiable points. It is clear from the Bar diagram that the P.T. values at S is higher than the P.T. values at R of V1 lead of normal ECGs whenever the opposite case holds for RVH ECGs. Also from Bar diagram we see that for V5 and V6 leads the difference between P.T. values at R and S point is small i.e. near to 1 for RVH ECG whereas these difference is large for normal ECGs. Also from tables of section 5 and 6 we see that the standard deviation at that point of our considerable leads indicate the P.T. values are consistent or not i.e. high and small values of standard deviation mean values are scattered and consistence respectively. Physically these mean the non-differentiable point of leads take near about same position for smaller value of standard deviation. Similarly the opposite case holds for higher value of standard deviation. So these are remarkable comparison of Normal ECG with Problematic ECG with fractional derivative techniques which is exactly new vision first adapted by us. As ECG is a rough curve so we will use Fractional Dimension, Hurst Exponent method etc. for our next consideration to give better solution. Thus by studying large number of ECG it is possible to find the a suitable range for the phase transition (P.T) values at the non-differentiable points that will help the doctors to determine the RVH conditions of patients. Also this type of study is not reported elsewhere. This method is a new method we are reporting for the first time- could be an aid for differential diagnostics in medical science.

## 9.Acknowledgement

Acknowledgments are to **Board of Research in Nuclear Science (BRNS), Department of Atomic Energy Government of India** for financial assistance received through BRNS research project no. 37(3)/14/46/2014-BRNS with BSC BRNS, title "Characterization of unreachable (Holderian) functions via Local Fractional Derivative and Deviation Function and to "**Indo-German Conference on Modelling, Simulation and Optimization in Applications" organized by Department of Mathematics, Bankura University** for giving me an opportunity to present this paper.